\documentstyle[prl,aps,multicol,epsfig]{revtex}
\renewcommand{\narrowtext}
{\begin{multicols}{2}\global\columnwidth20.5pc}
\renewcommand{\widetext}{\end{multicols}\global\columnwidth42.5pc}
\begin{document}
\newcommand{\be}{\begin{equation}}
\newcommand{\ee}{\end{equation}}
\newcommand{\bea}{\begin{eqnarray}}
\newcommand{\eea}{\end{eqnarray}}
\newcommand{\nt}{\narrowtext}\newcommand{\wt}{\widetext}

\title{Robust two-qubit quantum registers}
\author{I. A. Grigorenko and D. V. Khveshchenko}
\address{Department of Physics and Astronomy, University of North Carolina, Chapel Hill, NC 27599}
%\date{\today}
\maketitle

\begin{abstract}
We carry out a systematic analysis of a pair of coupled qubits,
each of which is subject to its own dissipative environment, and argue
that a combination of the inter-qubit couplings which provides for the lowest possible decoherence rates
corresponds to the incidence of a double spectral degeneracy in
the two-qubit system. We support this general argument by the results of
of an evolutionary genetic algorithm which can
also be used for optimizing time-dependent
processes (gates) and their sequences that implement various quantum computing protocols.
\end{abstract}
%\pacs{32.80.Qk}%\begin{multicols}{2}

The rapidly advancing field of quantum computing continues to bring about
a deeper understanding of the basic quantum physics alongside prospects of the 
ground-breaking potential technological applications. However, one of the
 main obstacles on the way of implementing quantum protocols has so far been
 and still remains a virtually unavoidable environmentally induced decoherence.

The list of the previously discussed decoherence suppression/avoidance techniques
includes such proposals as error correction\cite{error},
encoding into decoherence-free subspaces (DFS) \cite{dfs},
dynamical decoupling/recoupling schemes (e.g., "bang-bang" pulse sequences)\cite{bang},
and the use of the quantum Zeno effect \cite{zeno}.

However, none of the above recipes are entirely universal and
their actual implementation may be hindered by such factors as a
significant encoding overhead that puts a strain on the (still
scarcely available) quantum computing resources (error
correction), a stringent requirement of a completely symmetrical
qubits' coupling to the dissipative environment (decoherence-free
subspaces), a need for the frequency of the control pulses to be
well in excess of the environment's  bandwidth (dynamical
decoupling/recoupling) or an ability to perform a continuous high
precision measurement (quantum Zeno effect).
While being more feasible in some (most notably, liquid-state NMR and trapped-ion)
as compared to other quantum computing designs,
the implementation of the above approaches might be particularly challenging in solid-state architectures. 
It is for this reason that augmenting the above techniques with a many-body physics-conscious engineering of 
robust multi-qubit systems and a systematic approach to choosing the optimal (coherence-wise) values of their 
microscopic parameters appears highly desirable.

To this end, in the present work we explore yet another possibility of thwarting
decoherence by virtue of permanent (albeit tunable) inter-qubit couplings.
We find that, despite its being often thought of as a nuisance to be rid of,
the properly tuned qubits' "cross-talk" may indeed provide for an additional layer of
protection against decoherence.
Specifically, we focus on the problem of preserving an arbitrary
initial state ("quantum memory") of a basic two-qubit register during its idling period
between consecutive gate operations. Moreover, the optimization method employed in our work
can be further extended to the case of time-dependent two-qubit gates
which can alone suffice to perform universal quantum operations.

The previous analyses of the problem in question \cite{wilhelm,hanggi} have been largely limited to the 
symmetries of the underlying Hamiltonians and the parameter values corresponding to the presently available
 experimental setups \cite{phase,charge,devoret}.
As a result, they have not systematically addressed the issue of a
possible role of the inter-qubit couplings in reducing decoherence.
In contrast, our discussion pertains to a more general two-qubit
Hamiltonian 
\be 
{\hat{H}}=\sum_{a=x,y,z}[\sum_{i=1,2}{\hat\sigma}_a^{i}({B}^i_a+{h}_a^i(t)) +
J_{ab}{\hat\sigma}_a^{1}\otimes{\hat \sigma}_b^{2}],
\ee 
where each of the two
qubits described by an independent triplet of the Pauli matrices
${\hat \sigma}^i_a$ $(i=1,2)$ is subject to a local magnetic field
comprised of the constant ${\vec B}^i=(\Delta_i,0,\epsilon_i)$ and
fluctuating ${\vec h}^i(t)=(0,0,h_i(t))$ components, the latter
 representing two uncorrelated $(<h_1(t)h_2(0)>=0)$ dissipative reservoirs.
It is well known that the case of independent reservoirs present a significantly greater challenge 
than the greatly simplifying assumption of the highly correlated
ones where the standard DFS can be readily found \cite{dfs}.

As regards the symmetry of the qubits' interaction, we restrict
ourselves to the diagonal terms $\sigma_a^1\otimes\sigma^2_a$and
the associated parameters $J_{ab}=J_{a}\delta_{ab}$. Thus, we
exclude all the non-diagonal couplings between $\sigma_a^1$ and
$\sigma^2_b$ with $a\neq b$ which, albeit possible in principle,
do not normally occur in any of the known qubit designs.

As we argue below, the preferred symmetries of the inter-qubit
couplings can be found solely from the spectral analysis of the
noiseless part of the Hamiltonian (1). This suggests that our results should
be largely insensitive to the approximation used for treating the
effect of the noisy reservoirs.

To that end, we resort to the standard Bloch-Redfield (BR), i.e. a weak-coupling and
Markovian approximation, as does most of the previous work on the subject \cite{wilhelm}.
Although the BR approximation is known to become potentially unreliable in the (arguably, most
 important for quantum computing-related applications) short-time (as compared to the pertinent
 decoherence times) limit, the main advantage of the BR framework
is a relative physical transparency of its results.
Moreover, in light of the usual robustness of any symmetry-related conclusions,
such as those to be described below, we anticipate
that a more sophisticated analysis
(e.g., akin that of Refs.\cite{loss}) would fully corroborate the BR results.

In the standard singlet/triplet basis formed by the states:
$|\uparrow\uparrow>=(1,0, 0, 0)^T$, $(|\uparrow\downarrow>+
|\downarrow\uparrow>)/\sqrt{2}=(0, 1, 0, 0)^T $, $|\downarrow\downarrow>=(0, 0, 1, 0)^T$,
 and$(|\uparrow\downarrow>-|\downarrow\uparrow>)/\sqrt{2} = (0, 0, 0, 1)^T$,
the noiseless part of the Hamiltonian (1) takes the form
\be
\hat{H}_0=\left(\begin{array}{cccc} 
J_{z}+\epsilon&\Delta&J_{x}-J_{y}&-\Delta^{-}\\ 
\Delta &J_{y}-J_{z}+J_{x}&\Delta&\epsilon^{-}\\ 
J_{x}-J_{y}&\Delta&J_{z}-\epsilon&\Delta^{-}\\
-\Delta^{-}&\epsilon^{-}&\Delta^{-}&-J_{x}-J_{y}-J_{z}\\
\end{array} \right)
\ee
where $\Delta=(\Delta_1+\Delta_2)/\sqrt{2}$, $\Delta^{-}=(\Delta_1-\Delta_2)/\sqrt{2}$, 
 $\epsilon=\epsilon_1+\epsilon_2$ and $\epsilon^{-}=\epsilon_1-\epsilon_2$.

The BR equations for the reduced two-qubit (size $4\times4$) density matrix ${\hat \rho}(t)$ 
takes a particularly simple form in the basis of eigenstates of Eq.(2) \cite{wilhelm,weiss}
\be
\label{redfield}
{\dot{\rho}_{nm}(t)=-i \omega_{nm}\;\rho_{nm}(t)-\sum_{k,l}R_{nmkl}\; \rho_{kl}(t)}
\ee where $\omega_{nm}=(E_n-E_m)/\hbar$ are the transition frequencies,
and the partial decoherence rates
\be
\label{r_tensor}
R_{nmkl}=\delta_{lm}\sum_r\Lambda_{nrrk}+\delta_{nk}
\sum_r\Lambda^*_{lrrm}-\Lambda_{lmnk}-\Lambda^*_{knml}
\ee
are given by combinations of the matrix elements of the relaxation tensor \cite{weiss}
\be
\Lambda_{lmnk}=\frac{1}{8}S(\omega_{nk})\big[\sigma_{z,lm}^{1}
\sigma_{z,nk}^{1}+\sigma_{z,lm}^{2}\sigma_{z,nk}^{2}\big]+
\frac{i}{4\pi}{\cal P}\int_0^{\omega_c}S(\omega)d\omega\frac{\omega_{nk}}{\omega^2-
\omega^2_{nk}}\big[\sigma_{z,lm}^{1}\sigma_{z,nk}^{1}+\sigma_{z,lm}^{2}\sigma_{z,nk}^{2}\big]
\ee
determined by the products between the matrix elements $\sigma_{z,nk}^{i}$ computed in the
 eigenbasis of the noiseless Hamiltonian (2)
and the spectral density of the reservoirs
$S(\omega)=\int^\infty_0 dte^{i\omega t}<{\{}h_i(t),h_i(0){\}}>$$(i=1,2)$.

As in Refs.\cite{wilhelm},we choose $S(\omega)$ to be Ohmic, $S(\omega)=\alpha\omega\coth{\omega\over 2T}$,
 thus justifying the applicability
of the Fermi's Golden rule-based expression (5) (hereafter we use the units $\hbar=k_B=1$).

Obviously, the optimal choice of the Hamiltonian (2) would be the one that provides for
the lowest decoherence rates composed of the matrix elements (5).
The analysis of the expressions (5) reveals that their real parts tend to
decrease monotonically with the increasing length $J=|{\vec J}|$
of the vector ${\vec J}=(J_x,J_y,J_z)$. However, in all the
practically important cases (especially, in the before mentioned
Josephson junction-based designs \cite{phase,charge,devoret})
where the qubits emerge as effective (as opposed to genuine)
two-level systems, an unlimited increase of $J$ is not possible
without leaving the designated "qubit subspace" of the full
Hilbert space of the system. Besides, the coupling strength errors
that are likely to occur in any realistic setup tend to increase
with $J$ as well. Therefore, in what follows we fix the length of
the vector $\vec J$ and search for an optimal configuration of its
components,allowing for either sign of the latter.

Being proportional to the noise spectral density evaluated at frequencies corresponding
 to the transitions between different pairs of energy levels, the matrix elements (5), just 
like $S(\omega_{mn})$ itself, attain their minimum values 
(which are proportional to the reservoirs' temperature $T$) at $\omega_{mn}=0$.
Thus, one might expect that the relaxation processes will be quenched and
the decoherence rates will become suppressed, should 
some of the transition frequencies happen to vanish. 

Such a behavior would occur if any two of the energy levels of the 
Hamiltonian (2) became degenerate or, better yet, 
all the four levels of (2) became doubly (or even four-fold) degenerate. 
In this case, the lower pair of the degenerate levels 
could be viewed as an effective ("logical") qubit which is protected from decoherence
by an energy gap separating it from the upper pair of levels, resulting in 
an exponential suppression of the relaxation rates at low temperatures.  
Then, by encoding quantum information into this logical qubit's subspace, one can greatly 
suppress its decay, any residual decoherence being solely due to pure 
dephasing controled by $S(0)\sim\alpha T$. 

Next, we apply this general argument to the practically important case of identical qubits
($\Delta_1=\Delta_2=\Delta/2$) which are both 
tuned to the "co-resonance" point ($\epsilon_1=\epsilon_2=0$).

We note, in passing, that a special significance of this
parameter regime as a potentially most coherence-friendly
one is also evidenced by the experiments on the Josephson charge-phase qubits \cite{devoret}.

At the co-resonance point, the spectrum of Eq.(2) and the corresponding
eigenvectors are given by the following simpleexpressions:
$E_1=J_x+K, E_2=-(J_x+J_y+J_z), E_3=-J_x+J_y+J_z,E_4=J_x-K$ and
 $|\chi_1>=(1,(J_y-J_z+K)/\Delta,1,0)/{\sqrt{2K(K+J_y-J_z)}}$,
 $|\chi_2>=(0,0,0,1)$,$|\chi_3>=(-1,0,1,0)/{\sqrt 2}$,
$|\chi_4>=(1,(J_y-J_z-K)/\Delta,1,0)/{\sqrt {2K(K-J_y+J_z}}$,
where $K=\sqrt{(J_y-J_z)^2+2\Delta^2}$, respectively.

The only attainable incidence of double spectral degeneracy
($E_1=E_3, E_2=E_4$ or $E_1=E_2, E_3=E_4$) can occur for
$J_{y,z}>0$ or $J_{y,z}<0$, respectively. Resolving the above
conditions, we find that the conjectured optimal configuration of
the inter-qubit couplings must obey the equations \be
J^{opt}_x=0,~~ 2J^{opt}_y J^{opt}_z=\Delta^2 \ee By taking into
account the normalization of the vector $\vec J$ and Eq.(6) we
finally obtain 
$J^{opt}_{y}=\pm{1\over2}(\sqrt{J^2+\Delta^2}-\sqrt{J^2-\Delta^2})$,$~~J^{opt}_{z}=\pm{1\over
2}(\sqrt{J^2+\Delta^2}+\sqrt{J^2-\Delta^2})$. Notably, this result
shows that the conjectured optimal regime can only be achieved for
sufficiently strong couplings ($J>\Delta$).

Having elucidated the physical content of the expected
${\vec J}^{opt}$, we now confirm our predictions
by applying a direct optimization procedure based on a genetic algorithm \cite{holland}.
The latter starts out with random sets of parameters $\vec J$
which, in the language of Ref.\cite{holland}, constitute an
initial "population". In the course of the optimization procedure,
the worst configurations are discarded and certain recombination
procedures ("mutations") are performed on the rest of the
population. The new solutions ("offsprings") are selected
according to the values of the chosen fitness function. The
iterations continue until the set of parameters converges to a
stable solution representing the sought-after optimal
configuration ${\vec J}^{opt}$.

As a fitness function we choose such a customary quantifier of the register's performance as purity
$P(t)=\frac{1}{16}\sum_{j=1}^{16}Tr\{(\rho^j)^2(t)\}$
averaged over the solutions $\rho^j(t)$ of Eq.(3) with the initial conditions 
$\rho^j(0)=|\Psi_{in}^j><\Psi_{in}^j|$ given by 
the $16$ product states $\Psi_{in}^j=|\Psi_a>_1\otimes |\Psi_b>_2,(a,b=1,\dots,4$): 
$|\Psi_1>=|\downarrow>$,$|\Psi_2>=|\uparrow>$,
$|\Psi_3>=1/{\sqrt 2}(|\downarrow>+|\uparrow>)$,$|\Psi_4>=1/{\sqrt 2}(|\downarrow>+i|\uparrow>)$.

In Fig.1 we plot the purity decay rate $|dP(t)/dt|$ as a function
of $J_y$ and $J_z$ for $\epsilon_i=J_x=0$ and $\alpha=10^{-3}$. For
different values of $J$, the minima of of this two-parameter
function collapse onto a hyperbola in the $J_y-J_z$ plane which
appears to be described by Eq.(6) within the accuracy of our
numerical solution of the BR equations (3).

By varying the temperature, we observed that, consistent with our preliminary insight
into the mechanism of the decoherence suppression,
the effect of the interaction-induced
double spectral degeneracy appears to be
most pronounced at $T\ll\Delta$, while at
higher $T$ the landscape shown in Fig.1 flattens out,
thus making the optimization less effective.

We also observed that the use of an alternate performance measure,
fidelity $F=\frac{1}{16}\sum_{j=1}^{16}Tr\{\rho^j(t)\rho^j_0(t)\}$
where $\rho^j_0(t)$ represents the unitary evolution of an initial
state $\rho^j(0)$, yields the same results.

In order to further illustrate our findings, in Fig.2 we present
the behavior of a typical matrix element $R_{4123}$ as a function of a
 single parameter $w$ introduced by the relations: $J_x=0$,$J_y=0.25 \Delta w,
 J_z=1.98\Delta w$. It can be readily seen from Fig.2 that at the double 
degeneracy point $w=1$ which corresponds to $J=\sqrt{J_x^2+J_y^2+J_z^2}=2\Delta$ 
the matrix element $R_{4123}$ drops to its lowest value.

By contrast, in the non-degenerate case exemplified by the Ising configuration $J_x=J_y=0, J_z=2\Delta w$
 the above dramatic dropis now absent,
regardless of the value of $J=J_z$.
In Fig.3 we contrast the purity $P(t)$ computed for the 
optimal configuration with the results obtained in the Ising-, $XY$-, and 
Heisenberg-symmetrical, as well as non-interacting cases.
These results suggest a systematic way of constructing a class
of highly robust input states that undergo a slow decay governed by pure dephasing
 (no relaxation), as compared to the case of general encoding.

For $J_{y,z}>0$ a natural candidate for an orthogonal pair of such states is presented by 
the (anti)symmetrical combination
$|\chi^{stable}_{\pm}>=(|\chi_2>\pm |\chi_4>)/{\sqrt 2}$ of the lower pair of the
degenerate ($E_2=E_4$) energy levels.
The corresponding initial density matrix is characterized by the only non-zero entries 
$\rho_{22}(0)=\pm\rho_{24}(0)=\pm\rho_{42}(0)=\rho_{44}(0)=1/2$.

In Fig.4 we plot the decay rate of the purity $dP(t)/dt$ as a
function of the parameter $w$ for the initial state $|\chi^{stable}_+>$.
Note, that at the double degeneracy point ($w=1$) the purity decay rate is reduced by
several orders of magnitude.

The high robustness of a degenerate ground state facilitated by the
presence of the energy gap in our two-qubit system is somewhat reminiscent of 
the notion of "supercoherent" subspaces introduced in the abstract setting in 
Ref.\cite{bacon}. However, despite a certain resemblance between the two, 
our implementation of a robust logical qubit is different in a number of aspects. 

For one, the construction proposed in Ref.\cite{bacon} requires
at least four physical qubits which interact by virtue of the Hamiltonian
whose spectrum can be classified by the eigenvalues
of the total angular momentum of the system. 
Then the degenerate singlet ground state behaves as an "error detecting code"
that can only loose its coherence by absorbing energy from the reservoirs,
the rate of which process appears to be exponentially suppressed for $T\to 0$.

In contrast, our logical qubit consists of only two physical qubits
interacting via a generic (not necessarily spin-rotationally invariant) Hamiltonian, 
and its decoherence rate is limited by pure dephasing whose contribution remains 
linear in $T$ even at the lowest temperatures. However, albeit not providing 
an equally strong protection for the logical qubit itself, 
in our case the double degeneracy of the two-qubit spectrum does improve the gate 
performance for general encoding, as manifested by the gate 
characteristics averaged over different initial states.

One can also identify the initial states whose decay can only get
worse upon increasing the qubit coupling. These fragile states
belong to the subspace spanned by the complementary pair of
eigenstates $|\chi_1>, |\chi_3>$.

Lastly, a word of caution is in order. Conceivably, any incidence of degeneracy
may invalidate the use of the Golden rule-based Eq.(5) whose (sufficient) conditions of 
applicability requires that $max|R_{mnkl}|\ll min|\omega_{mn}|$.
However, we found that our numerical solution remains stable even in a close vicinity of 
the double degeneracy point.

The optimal configuration ${\vec J}^{opt}$
shows no abrupt changes upon varying the strength of the
coupling $\alpha$ from its lowest values that do comply with the above sufficient conditions
all the way down to the parameter range where $min|\omega_{mn}|\ll max|R_{mnkl}|\ll max|\omega_{mn}|$.

To summarize, in the present work we demonstrated that the properly chosen
permanent inter-qubit couplings can provide a new means of protecting quantum
registers against decoherence. While being relatively unexplored in the mainstream quantum 
information proposals, a similar idea has been in the focus of a number of scenarios exploring 
the possibility of a natural emergenceof logical qubits and the conditions facilitating fault-tolerant 
computations in strongly correlated spin systems.

In the existing proposals \cite{kitaev} the prototype logical
qubits are envisioned as topological ground and/or quasi-particle
bulk/edge states of rather exotic chiral spin liquids. Thus,
despite their enjoying an exceptionally high degree of coherence,
the(topo)logical qubits require an enormous overhead in encoding,
since creating only a handful of such qubits takes a
macroscopically large number of interacting physical ones.
Besides, the nearly perfect isolation of (topo)logical qubits from
the environment can also make initialization of and read-out from
such qubits rather challenging.
Nonetheless, our results show that a somewhat more modest idea of
augmenting the other decoherence-suppression techniques with
appropriate tailoring of the inter-qubit couplings  might still
result in a substantial improvement of the quantum register's performance.

The numerical optimization method employed in this work can be
further extended to the case of time-dependent two-qubit gates as
well as beyond the Bloch-Redfield approximation and/or the
assumption of the Ohmic dissipative environments.

As regards the gate optimization, we find
that in many cases the best performance of a two-qubit register
can be achieved with a rectangular pulse $J_a(t)=J_a\Theta(T-t)\Theta(t)$
where the choice of the coupling parameters $J_a$ facilitates
the degeneracy of the lowest pair of energy levels. These results will
be presented elsewhere \cite{future}.

The rapid pace of the technological progress in solid-state
quantum computing gives one a hope that the specific prescriptions
towards building robust qubits and their assemblies discussed in
this work can be implemented in future devices. In this regard,
very promising appear to be the phase \cite{phase},
charge\cite{charge} and charge-phase \cite{devoret}
superconducting qubit architectures where, in principle, one can
achieve any desired symmetry of the interaction terms in Eq.(1) by
merely tuning the capacitive and inductive couplings between
different Josephson junctions which implement the physical qubits.

The authors acknowledge valuable discussions with J.E. Mooij and M. Devoret.

This research was supported by ARO under Contract DAAD19-02-1-0049.

\begin{figure}
\begin{center}
\includegraphics[height=0.5\textwidth,angle=0]{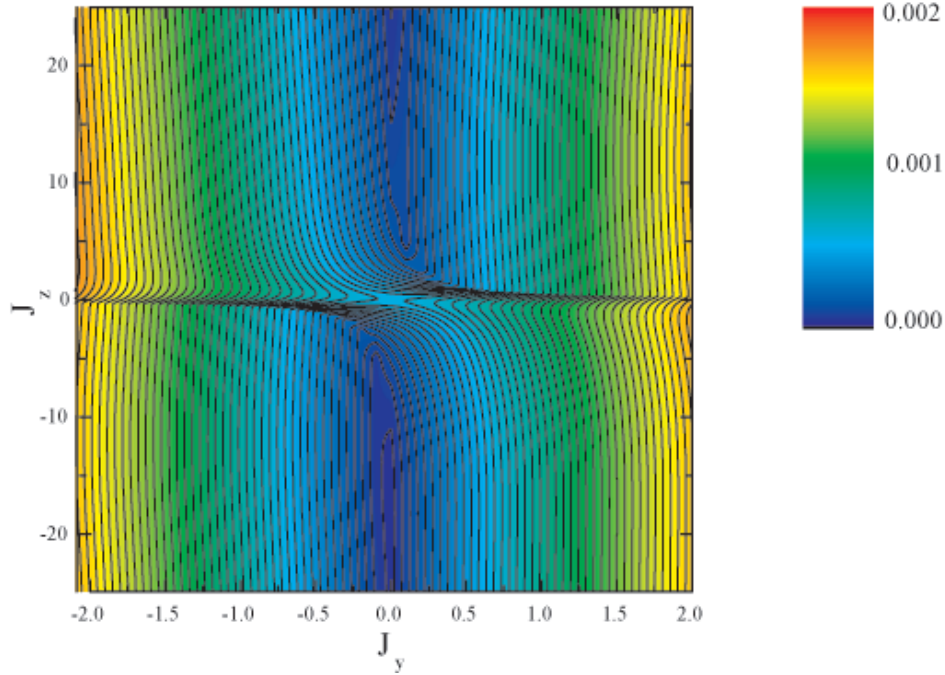}\vspace{0.3cm}
\caption {Purity decay rate $dP/dt$ as a function of $J_y$ and $J_z$ for
 $J_x=0$ and $T=10^{-5}\Delta$.}
\end{center}
\end{figure}
\begin{figure}
\begin{center}
\includegraphics[height=0.5\textwidth,angle=-90]{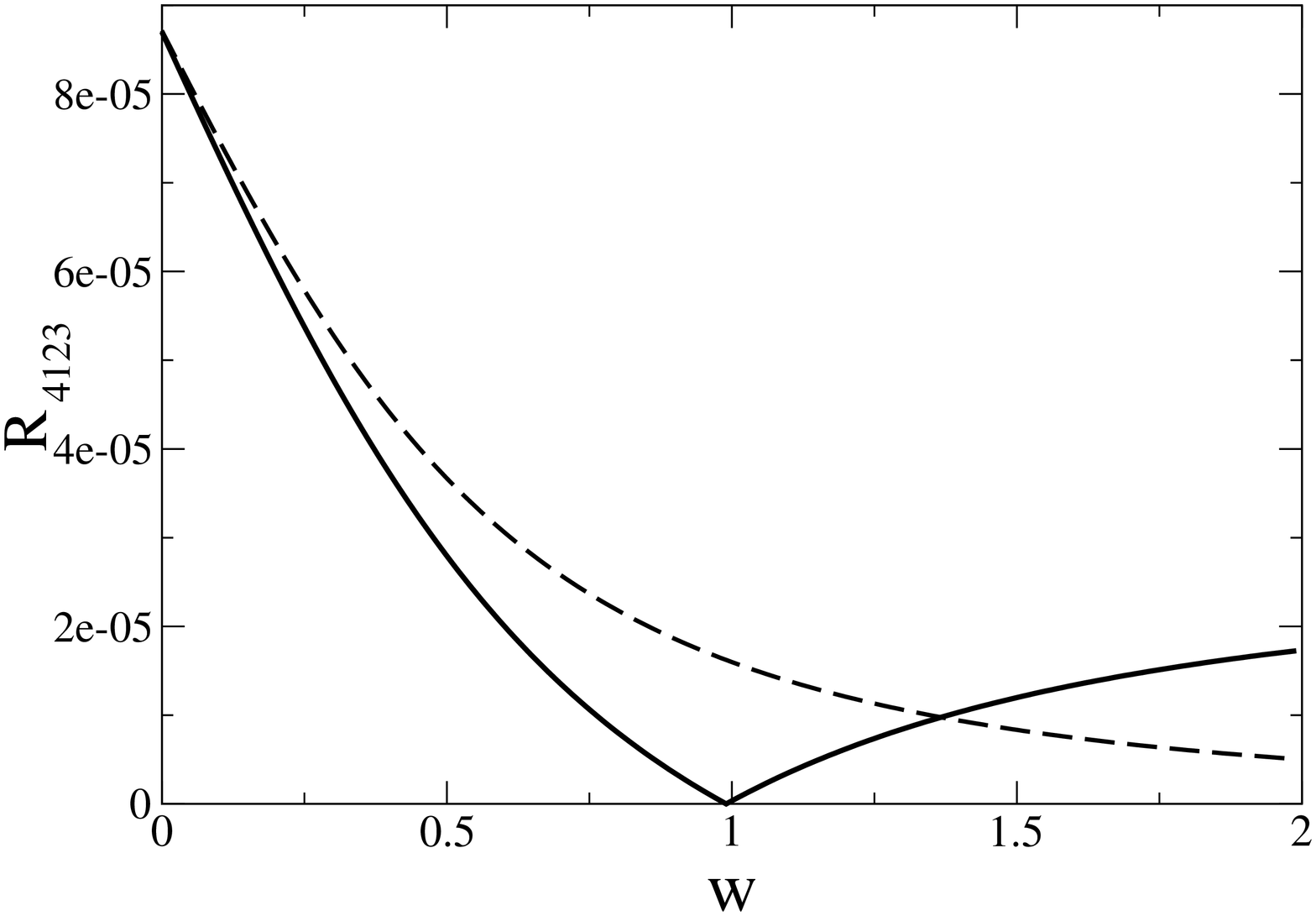}\vspace{0.3cm}
\caption{Partial dephasing rates exemplified by a typical matrix
element $R_{4123}$ as a function of the parameter $w$ computed for
$T=10^{-5}\Delta$, ${\vec J}=(0, 0.25\Delta w, 1.98\Delta w)$
(solid line) and ${\vec J}=(0, 0, 2 \Delta w)$ (dashed
line).}
\end{center}
\end{figure}
\begin{figure}
\begin{center}
\includegraphics[height=0.5\textwidth,angle=-90]{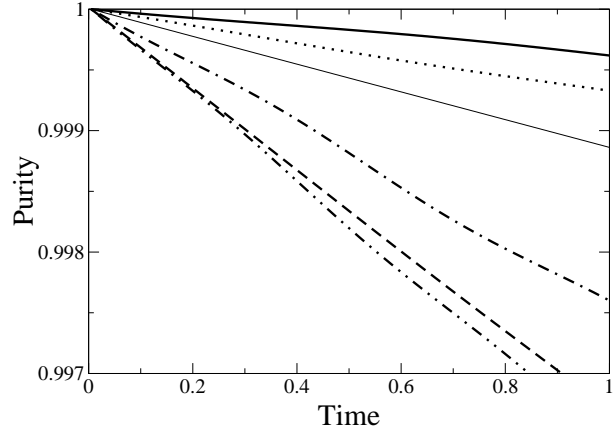}\vspace{0.3cm}
\caption {Purity as a function of time for zero coupling (${\vec
J}=0$, thin solid line); Ising (${\vec J}=(J, 0, 0)$, dotted line
and ${\vec J}=(0, J, 0)$, dashed line), $XY$ (${\vec J}=(J/{\sqrt
2}, J/{\sqrt 2}, 0)$, dash-double dotted line), Heisenberg (${\vec
J}=(J/{\sqrt 3}, J/{\sqrt 3}, J/{\sqrt 3})$,dash-dotted line), and
the optimal configuration(${\vec J}={\vec J}^{opt}$, solid line).
Here $J=2\Delta$ and
$T=10^{-7}\Delta$}
\end{center}
\end{figure}

\begin{figure}
\begin{center}
\includegraphics[height=0.5\textwidth,angle=-90]{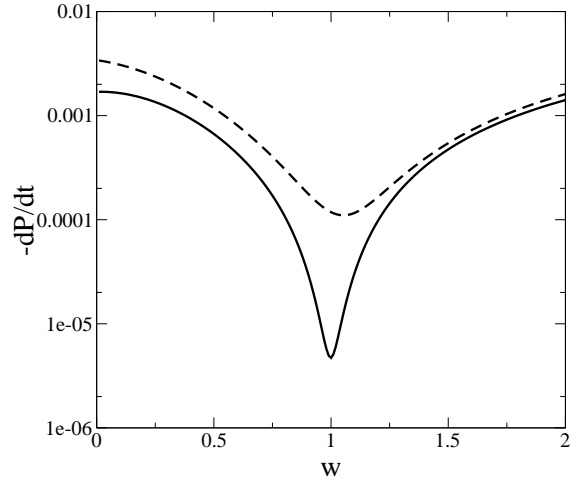}\vspace{0.3cm}
\caption {Purity decay rate $dP/dt$ as a function of the parameter
$w$ for ${\vec J}=(0, 0.25\Delta w, 1.98\Delta w)$for
$T=10^{-5}\Delta$ (solid line) and $T=\Delta$ (dashed
line).}
\end{center}
\end{figure}
\end{document}